# Low-cost modular devices for on-road vehicle detection and characterisation

Jose-Luis Poza-Lujan[1] · Pedro Uribe-Chavert[2] · Juan-Luis Posadas-Yagüe[1]

## Abstract

Detecting and characterising vehicles is one of the purposes of embedded systems used in intelligent environments. An analysis of a vehicle's characteristics can reveal inappropriate or dangerous behaviour. This detection makes it possible to sanction or notify emergency services to take early and practical actions. Vehicle detection and characterisation systems employ complex sensors such as video cameras, especially in urban environments. These sensors provide high precision and performance, although the price and computational requirements are proportional to their accuracy. These sensors offer high accuracy, but the price and computational requirements are directly proportional to their performance. This article introduces a system based on modular devices that is economical and has a low computational cost. These devices use ultrasonic sensors to detect the speed and length of vehicles. The measurement accuracy is improved through the collaboration of the device modules. The experiments were performed using multiple modules oriented to different angles. This module is coupled with another specifically designed to detect distance using previous modules' speed and length data. The collaboration between different modules reduces the speed relative error ranges from 1 to 5%, depending on the angle configuration used in the modules.

Jose-Luis Poza-Lujan, Pedro Uribe-Chavert and Juan-Luis Posadas-Yagüe have authors contributed equally to this work.

✉ Jose-Luis Poza-Lujan
  jopolu@upv.es

  Pedro Uribe-Chavert
  pedurcha@doctor.upv.es

  Juan-Luis Posadas-Yagüe
  jposadas@upv.es

1 Research Institute of Industrial Computing and Automatics (ai2)., Universitat Politècnica de València (UPV), Camino de vera sn, 46980 Valencia, Spain

2 Doctoral School, Universitat Polièncica de València (UPV), Camino de vera sn, 46980 Valencia, Spain

## 1 Introduction

Cyber-physical systems (CPS) allow the seamless integration of computing and physical objects in our daily lives [1]. This integration transforms how people interact with the physical world [2]. In recent years, the rapid development of CPS has driven significant innovations in a wide range of application domains such as intelligent manufacturing, smart homes and communities, autonomous driving or transportation support. CPS requires a distributed and decoupled infrastructure with high adaptability, scalability, strength, and security. This distribution of elements and functions becomes especially critical when managing real-time systems actions must be completed within particular time frames [3].

Among the different areas of application of CPS, highlights the intelligent management of transport (ITS) [4] of both people and goods. ITS achieves traffic management to detect problematic situations, such as vehicles that circulate improperly or see accidents [5]. These aspects have led to various fields of research on how the elements of intelligent transport systems interact between them [6]. With intelligent transport systems, vehicles can interact with each other (V2V) or with the transport infrastructure (V2I). Likewise, the infrastructure can communicate its elements with each other (I2I). In the ITS, it is necessary to characterise the traffic profile to optimise the road. Consequently, knowing the traffic profile is essential as predictive systems require training and updating based on accurate data [7]. A system that uses the size and speed of vehicles can predict traffic jams, crowds, and infractions, or even adapt lighting to the traffic on the road.

There are many methods of detecting vehicles on roads [8]. The most efficient methods use complex devices, such as cameras [9] or even drones [10]. These systems are very suitable in environments with a high density of vehicles with a high probability of incidents, such as accidents. They are also suitable in emergencies, where low-cost sensors cannot supply the global vision needed to vehicle surveillance [11]. However, the price of these systems or availability are points that only recommend their use in well-lit urban environments or particular situations. Also, these systems based on images have a high computational cost, making them more complex computer-based devices. Last but not least, if the images are sent to processing nodes, the cost of communications also increases. These devices can be tempting for vandalism, in addition to not having the availability of high processing or communicating capacity [12]. Consequently, in the case of roads isolated from urban environments, detection systems that use high-cost elements or high-requirements elements must be replaced by lower-cost ones. Low-cost devices can be placed closer to the road, for example, on guardrails or traffic signs.

If on-road vehicle detection systems use devices located close to the vehicles, it is necessary for the devices to use sensors with a short operating range. Consequently, imply that on-road systems usually require several devices to spread out. As a result, information is scattered and cooperation between devices [13] and between vehicles [14] becomes critical. In these cases, systems are no longer oriented to precisely recognize vehicles but are closer to recognising their behaviour. Some detectable behaviours are the control of the trajectories [15], the detection of dangerous situations [16] or the detection of infractions [17]. The system should be distributed with devices that can be adjusted depending on where they are placed [18].

As far as the low-cost distance sensors of the system devices are concerned, there is some consensus on the use of ultrasound technology because of the better accuracy it provides related to the sensor price [19]. The devices can be mounted in different positions for vehicle detection: overhead mount, side-top mount, and horizontal mount [20]. The location selection depends on the availability of a suitable installation, such as bridges for the overhead mount, poles for the side-top mount, or elements close to the ground for the horizontal mount. A distance sensor can provide vehicle characteristics such as length or speed depending on position and orientation. In [21], a study shows how it is possible to obtain the speed of a vehicle with ultrasonic sensors, where the accuracy of the result depends on the vehicle's speed.

To obtain more accurate data, it is better to have several sensors. In [22], an ultrasonic array is presented but mounted along the streets to simultaneously detect and track several vehicles. In this case, sensors are only used to recognise the presence of vehicles and centrally fuse the data to characterise the traffic. However, with multiple similar sensors, more information can be obtained. For example, in [23], an array of ultrasonic sensors mounted on a vehicle is used to recognise the vehicle environment. In this way, the vehicle can characterise its contour, and results indicate that it is possible to obtain a vehicle speed with accuracy from $\pm 0.2$ to $\pm 1$ m/s. This characterisation can be used to recognise traffic evolution. In [24], sensors are placed along streets, or urban roads, to recognise traffic density. In this case, only the presence or not of a vehicle is measured, without characterising aspects such as speed or length that would allow to predict traffic better and manage, for example, traffic lights in the case of urban environments. The use of several sensors in the same device enriches the information obtained. For example, in [25], two sensors characterise the length and type of vehicle but not the speed. The results are pretty promising and indicate that the simultaneous use of distance sensors provides more detailed vehicle information.

This article presents a system based on modular devices. The use of modules in devices is an aspect widely used in connected embedded systems [26]. The detection modules are based on a low-cost ultrasonic sensor. These modules obtain the speed and length from a simple linear computational cost algorithm. The modules cooperate between them to improve the precision of the vehicle's length calculation. Unlike the previously presented systems, the novelty of the device shown in the article lies in the simultaneous characterisation of the length, as in [25], and the vehicle speed, as in [23]. Another novelty is to include the reinforcement of the data obtained in the same device or by collaborating with neighbours, as presented in [24].

The article is structured as follows. The Sect. 2, material and methods, begins by describing how it can characterise a vehicle with a distance sensor. Below, the system is shown roughly to describe the method of detection and characterisation of vehicles. Finally, the algorithms implemented in each module to filter, detect vehicle parts and characterise vehicle length and speed are presented. Section 3 explains the validation process for the proposed method. Specifically, an Arduino prototype was used to validate the method. The section ends with a description of the experimental device developed. The first experiment evaluates different sensor orientation angles to detect speed, length, and errors. Then a device with two modules dedicated to measuring speed and distance was evaluated. The following experiment was performed in a python simulator by modifying an existing traffic simulator. In this simulator, five modules were simulated with five different angles. When the simulation ends, the data obtained are processed to obtain the speed and length of each angle. It has been verified how the collaboration between the two modules improves the accuracy of vehicle length detection. Finally, the Sect. 4 show the conclusions of the paper.

## 2 Materials and methods

### 2.1 Ultrasonic sensors for vehicle detection

As mentioned above, the distance sensor can be theoretically modelled as a beam that cuts the vehicle and allows devices to obtain an approximate distance measure. The distance sequence

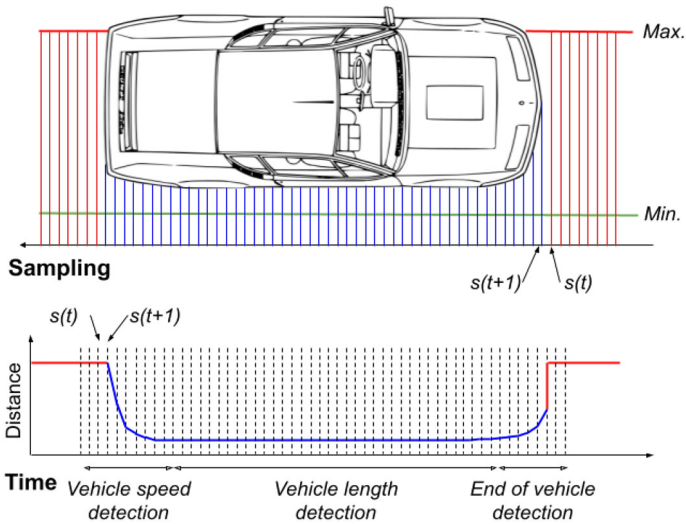

**Fig. 1** Measurements and distance patterns are obtained when the ultrasonic sensor is perpendicular (90°) concerning the axis of the road. The result is a profile where it is relatively easy to detect patterns of the entrance and exit of a vehicle in the operating range of the sensor

obtained by the sampling over time generates a vehicle profile. Figure 1 shows the sampling along the vehicle passed with a sensor at 90° on the axis of the road. The profile obtained from measurements throughout said sampling is also shown.

It is possible to detect patterns from the profile measurement. These patterns can detect the start and the end of the vehicle. On the detection of the vehicle, depending on its morphology, some beams could not coincide with the vehicle's frontal, which it would make difficult to calculate a speed approach. So, a sensor positioned at 90° from the axis of the road will not be able to obtain an accurate sample of the speed, especially if we consider that, at this angle, the vehicle is very similar to a rectangle. To improve speed measurement accuracy, placing a more significant number of rays on the front of the vehicle is recommended. To do this, starting from the fact that the sensor must be on one of the road edges, it will have larger samples belonging to the front if it turns. Figure 2 shows how the sampling would be in the case of a sensor rotated 45° concerning the axis of the road.

In this case, the number of samples belonging to the front of the vehicle is more significant. With a significant number of samples, it is possible to calculate the speed more accurately. From the speed detection, the length calculation is immediate.

### 2.2 Vehicle detection and characterisation

Many measurements from the ultrasonic sensor in a row imply various signal processing must be applied. [27]. This process involves successive treatments of the signal obtained shown in Figs. 1 and 2. In the case of the system presented, the signal processes involved are shown in Fig. 3.

As these are three different types of processing, it is possible to implement them separately. This process separation is proper when the system infrastructure has certain limitations

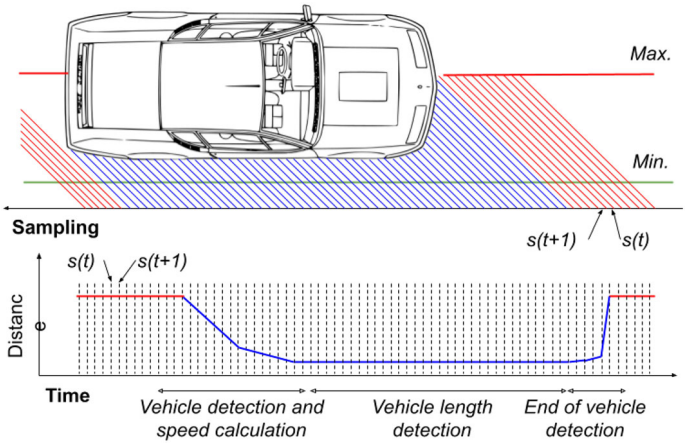

**Fig. 2** Measurements and distance patterns are obtained when the ultrasonic sensor is perpendicular (45°) concerning the axis of the road. In this case, the resulting pattern is different from that of the device positioned perpendicular (Fig. 1). By comparing these differences, it is possible to adjust the values obtained

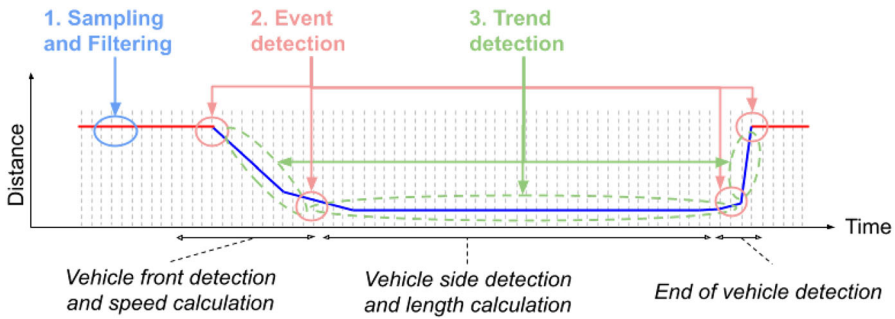

**Fig. 3** Processing of the signal obtained by successive sensor measurements. First, the ultrasonic signals are sampled and filtered (1). Next, a trend change event detection is implemented to detect the vehicle's front or side (2). Finally, the signal trend allows making a hypothesis about the speed or length (3)

in processing and communications. The following will explain how these processes are distributed in different phases to be embedded in a system.

Figure 4 shows the blocks and the functioning of the sensor module to calculate and provide speed and length.

The first phase of sampling and filtering of the sensors' measurements aims to provide a signal, a sequence of measurements, without outliers or peaks and softens the measurement variations. The event detection module must process this signal; specifically, the events that must be detected are the trend changes. These trend changes determine the main events. The first event is the start front of a vehicle. The front appears when the signal goes from a flat to a negative slope. The second event determines when the front of the vehicle ends and starts on the vehicle's side, which means that the signal passes from a negative slope to a zero slope. Finally, the last event determines when the signal slope becomes positive and the slope returns to zero. Therefore, the events, together with the trend changes, allow for calculating the speed and length of the vehicle.

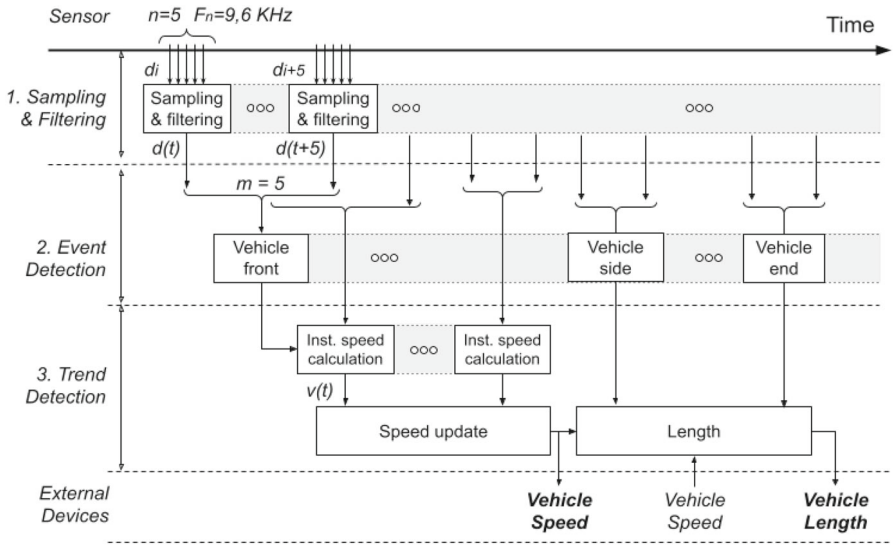

**Fig. 4** Block diagram of module processing. The speed and length calculation can be enriched with data from external devices. It is possible to characterise the vehicle more accurately

### 2.2.1 Sampling and filtering

The sampling phase runs continuously, reading from the sensor five distance measurements at a frequency of 50 Hz. This frequency assumes a sample period of approximately every 20 ms. Due to the ultrasonic speed being about 340 ms per second, the sampling period avoids a lot of echoes from previous samples. The ultrasonic sound is pulsed above 40 kHz, so it is not audible to the human ear. The first filtering process consists of detecting the outliers, or values out of range, exceeding the maximum and minimum values the sensor can provide. The second filtering process is to detect peaks in the sequence of measures. In this case, a peak is considered when a value, into the range, that exceeds the previous and subsequent values by a percentage. The percentage, in our case, has been determined experimentally at 15% because minor variations occur depending on the material's characteristics that reflect the sensor signal. Longer duration spikes are filtered out as they could indicate an item protruding from the detected vehicle. The sample is discarded if the number of abnormal values in the array exceeds the number of valid values.

In the case of ultrasound sensors, the signal obtained from the sensor has low noise, and sensor fusion is performed in late stages, so it is not necessary to use advanced filtering. Consequently, the sampling is based on a sliding window using a circular buffer, so it is more efficient in processing time to use an exponential moving average (EMA) filter widely used in ultrasound sensors [28]. Equation 1 shows the formula to calculate EMA.

$$A_n = \alpha \cdot M + (1 - \alpha) \cdot A_{n-1} \qquad (1)$$

In Eq. 1, $A_n$ is the filtered value, $A_{n-1}$ is the previous filtered value, $M$ is the sampled value of the signal to be filtered, and $\alpha$ is a factor between 0 and 1. The $A_{n-1}$ factor provides a smoothing that depends directly on the value measured. The $\alpha$ factor is a condition for the exponential filter, and its variable values range between 0 and 1. In the case of setting $\alpha = 1$, an unfiltered signal is obtained; an $\alpha = 0$ returns a filtered value equal to 0. Decreasing the value of α increases the smoothness of the signal, making it more stable and less susceptible to noise, but causes an increase in the system's response time, delaying the signal with respect to the original. The most common values range from 0.2 to 0.6. They were 0.75, the value that provided the best signal to be able to be processed later.

There are different methods to perform efficient signal smoothing, peak reduction, and sensor fusion, such as the Kalman filter [29] or the Weighted outlier-robust Kalman (WRKF) [30]. However, EMA has advantages in the coding since it uses simple arithmetic instructions and stores only the previous filtered value in memory. Consequently, the EMA filter quickly provides excellent results when α is tuned adequately and uses few computational and spatial resources.

### 2.2.2 Event detection

Recognising that the vehicle's front, side, and the back part has entered the sensor range involves event detection.

However, since the same module performs the acquisition and filtering, computational resources are scarce, so a Cumulative Sum control chart (CUSUM) Detector [31] is used in the presented system. Cusum is an algorithm of sequential analysis that allows abrupt changes in the trend of continuous signals. It allows for detecting the variations early by simple calculations, which facilitates coding the algorithm in systems with few computational or memory resources, such as embedded systems based on low-cost processors. The algorithm is based on the fact that a signal is stable if the value of a measurement $A_n$ is similar to the average of the previous values $\overline{A_{[n-1,e]}}$ (being $e$ the first measure in the current trend) with a $z$ variation on the standard deviation $\sigma$ (Eq. 2). The variation Z is critical since it is considered that a value greater than 3 implies that the measure measured $A_n$ has been diverted from the trend and, therefore, we have a change of trend.

$$A_n = \overline{A_{[n-1,e]}} + z \cdot \sigma \quad (2)$$

To detect the change in the trend the value of Z is calculated in each new measure $A_n$. This value is immediate from Eq. 3.

$$z = (A_n - \overline{A_{[n-1,e]}})/\sigma \quad (3)$$

One of CUSUM's weaknesses is that the detection of changes depends on the threshold in which it is considered that $z$ indicates a change in trend. It is common for a value of $z = 3$ to be considered sufficient for a noticeable change. However, to increase the precision of the point where the change in trend is detected, a second-derivative test [32] is made. The second derivative makes it possible to detect peaks and troughs, and thus the trend change of the signal, more accurately than CUSUM. However, to achieve the forecast, it is necessary to have a large amount of data preceding and consequent to the event that determines a trend change. This amount of data is collected as the vehicle moves, which is used to detect the signal trend, in our case, speed and length. It is, therefore, possible to adjust the points at which the vehicle's front, side or back is detected as more measurements are obtained. How the speed and length are obtained using the first and second derivatives of the signal is explained below.

### 2.2.3 Trend detection

After recognising the relevant points of the vehicle, the speed and length are calculated from the trend of the data obtained by the front, the side, of the back of the vehicle. The trend

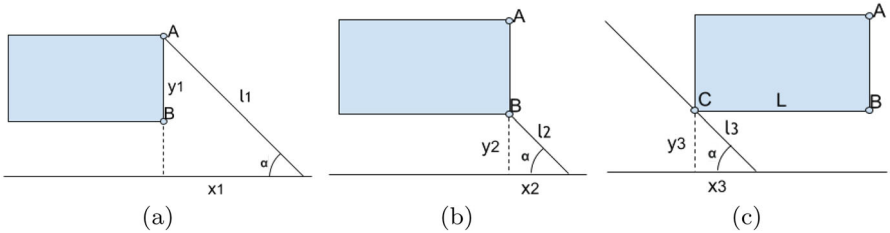

**Fig. 5** The three points used to calculate the speed and length of the vehicle. Figures show the critical points determined by the trend change events and the parameters used to calculate the vehicle speed and length

change events in the signal define the points that will determine whether the vehicle's front or side is being detected (Fig. 5).

There are many methods of trend detection. Those based on statistical analysis, such as linear regression [33] or correlation methods, such as the Likelihood Ratio Test [34], require the entire sample to perform the detection. In our case, to obtain an approximation of the speed as soon as possible, a derivative test [35] is used with the first derivative with reinforcement of the second derivative, performing the same derivative test.

The speed is calculated from the trend of the measures obtained between points A and B in Fig. 5. The Eq. 4 continuously uses the derivative of the length $l$ as measurements are obtained. In the equation, $\alpha$ is the angle of the sensor oriented

$$v(t) = \frac{d \cdot}{dt} l(t) \cdot \cos \alpha \qquad (4)$$

Length is calculated from the trend of the data obtained between point B and point C in Fig. 5. The formula 5 is used to obtain length and uses

$$L_{B-C} = \frac{v}{\Delta t_{B-C}} \qquad (5)$$

Only speed and length with one accuracy can be obtained using a single angle, as shown in Fig. 5. Different angles obtain different accuracy. The highest accuracy in speed would be given by a sensor with a beam parallel to the road axis, with a $\alpha = 0$. This angle would not be able to measure longitude, besides the difficulty of integrating the sensor in front of a vehicle. A sensor would provide the highest accuracy in length with a $\alpha = 90$, i.e. perpendicular to the road axis. However, such an angle would not be able to measure speed but would give high accuracy in length if one has an initial value of speed as close as possible.

From different values of $\alpha$, additional speed and length accuracy can be obtained. To have different angles, different sensors must be available. Therefore, the concept of a single sensor and the combination of different sensors, with different orientations, will be tested in the experiments. The vehicle speed verification and length calculation phase of the vehicle characterisation consist of collecting the information corresponding to speed and length from various modules. These values are dependants, as seen in Figs. 1 and 2 of the sensor angle. An experimental characterisation, shown in the corresponding section, allows for knowing the absolute error associated with the angle and speed of the vehicle.

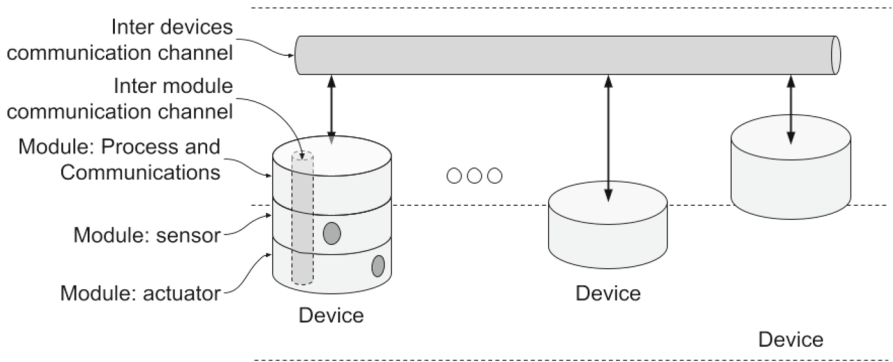

**Fig. 6** Overview of the system presented. Devices are made of modules. Modules can be dedicated to sense, act, compute and communicate data. Sensor modules provide single values, and computing modules verify the diverse values and increase the accuracy of the vehicle characterisation

## 3 Experiments and results

Different experiments have been carried out to validate the presented method and are described in this section. The first sub-section shows the concept of the system to have experimented with, where the devices contain modules. Each module has a single sensor oriented with an angle, which can share its information with other modules with different angles. The following subsection describes the technical characteristics of one module implemented. This module allows us to validate the method and characterise the sensor's signals to simulate more modules. The third grant describes the scenarios experienced. The first of the scenarios consists of a single module tested with a different angle. This scenario allows us to characterise the behaviour of the sensor at each angle. The accuracy is limited when only one module is used to calculate speed and length. Therefore, the third scenario is to combine more than two modules to test if having many angles improves the calculation of both speed and length. One of the research objectives this work contributes is to have an extensive network of sensors that monitor long distances on streets and highways. When testing the performance of many modules together, it is desirable to start by simulating the system before implementing the physical system. For this reason, a simulator has been modified to include the This third scenario, and future ones with more modules, are presented with the simulated data.

### 3.1 System proposed

It is proposed to use a distributed system based on connected modular devices to carry out the measurements and adjust their precision. Figure 6 shows the system used. The modules are defined as the minimum unit of the system and contain elements that connect with the physical environment, sensors, and actuators. In the case of the presented system, only distance sensors are used. A device comprises a series of modules connected by an internal communication channel. The devices communicate with each other through a proximity channel, such as bluetooth or power-line [36].

In the case of the presented system, a device can merge the information from various modules. As shown in Fig. 6, there may be devices with only sensors or actuators that do not improve the collected data, as well as devices without sensors or actuators that are dedicated to the more advanced processing. Once the vehicle is detected, the module proceeds to calculate vehicle speed, detecting the distances of the front and the side of the vehicle. Finally, when the sensor returns to provide the maximum distance, it is considered that the vehicle's transition has already finished.

## 3.2 Experimental device

The experimental setup is based on identical ultrasound modules, presented in [37]. Modules have been built from the JSN-SR04T 2.0. ultrasound sensor module. This sensor is waterproof and widely and widely used both in automotive and for liquid tank level measurement. This sensor is widely used due to its adaptability to outdoor use and accuracy [38]. The sensor has a detection range of 0.25 m at 4.5 m. This range makes it very suitable for covering lane vehicle profiles. The resolution of the distance is 0.005 m, so it allows relevant variations so that the detection algorithm can work efficiently. The sensor sampling rate has been set to 9.6 KHz. The sensor has been connected to an Arduino 'Nano' which, in turn, communicates with the device's modules via inter-integrated circuit (I2C). This channel, designed by Philips [39], allows serial communication between a master and several slaves at speeds between 100 Kbits/s and 3.4 Mbits/s. It is a channel widely used in embedded systems due to its simplicity in management. The connection between various devices is made through a Bluetooth 4.0 HM-10 interface to be able to make bi-directional connections with nearby devices with low energy cost [40].

For the experiments, a vehicle with a length of 3.7 ms was used. The vehicle has been passed through the measuring module ten times for angles of 30°, 45°, and 90°. For each angle, two speeds have been passed: 10 m/s (36 km/h) and 20 m/s (72 Km/h). These speeds are selected for ease of being measured during testing quality radar. In this way, those steps of the vehicle in which the speed differs from that established for the experiment can be discarded.

## 3.3 Experimental scenarios

Two scenarios have been taken into account, one for validating the characterisation and the other to measure optimisation obtained when two modules collaborate (Fig. 7b. In the first of the scenarios, a device with only one module is characterised. In this phase, the module is regulated successively and configured with angles of 30° and 45° degrees. From 45 degrees, the relative speed error is so high that it is difficult to use it to calculate length with an admissible accuracy. With these modules, the precision in the calculation of the speed and the accuracy in the computation of the length from the calculated speed is evaluated. In other words, the entire measurement process described in Fig. 4 is carried out in the same device.

The second of the scenarios consists of including two modules in a single device, In this case, a 30° module is used to measure the speed, and a 90° module is used to, using the 30° module speed, calculate the length with the precision of 90°. In this second scenario, a device can improve precision locally.

### 3.3.1 One-module device

The first experiment consists of characterising a single module using three different angles. In this case, 30°, 45°, and 90° have been chosen. With an angle of 30 degrees, the module has an orientation as parallel as possible to the road, so it is a reasonable estimate of the most precise speed possible to obtain. However, the calculation of the length is not particularly precise.

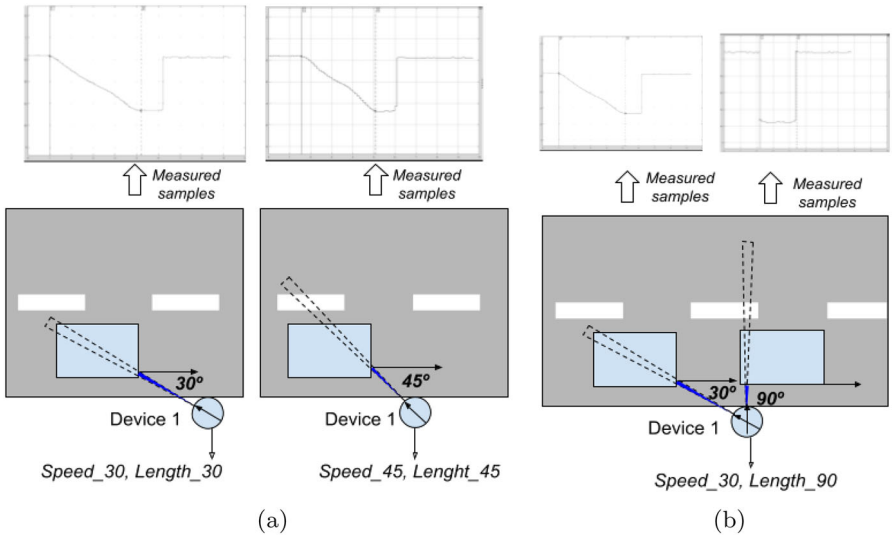

**Fig. 7** Experimental scenarios tested. Scenario 1 **a** consists of the test in the same device with two different angles. Scenario 2 **b** consists of two modules collaborating to obtain an accurate measure

**Table 1** Experimental results of the speed measures obtained by a module as a function of the orientation angle

|  | 30° | | 45° | |
| --- | --- | --- | --- | --- |
|  | 10 m/s | 20 m/s | 10 m/s | 20 m/s |
| AVG (m/s) | 10.55 | 21.12 | 12.92 | 24.04 |
| STD (m/s) | 1.01 | 3.02 | 1.18 | 2.40 |
| Abs. error (m/s) | 0.58 | 1.74 | 1.78 | 1.38 |
| Rel. error (%) | 5.79 | 8.71 | 17.77 | 6.91 |

With the 45° angle, the accuracy of the speed decreases, and length accuracy increases. A single 90° module is not contemplated in the measurements due to its inability to measure the speed, so the length cannot be calculated without this angle.

Concerning the length, since a calculated speed is already available, this depends on the accuracy. Table 2 shows the results of the calculated lengths.

### 3.3.2 Two-module device

Once the accuracy of the speed and the length are obtained by sensors oriented at different angles, we experiment with how the calculated rate can be used to improve the length of the vehicle by adding a 90° module. For this experiment, a 30° module is available in the same device since it provides more precise information on speed. The two modules communicate over the I2C bus, with a third module that acts as a master and, in addition, manages the wireless communications with other devices for other configurations. In this way, the processing loads of each sensor module are not altered by extra processing in the measurement optimisation phase. Final results are shown on chart 3

As seen in the results, the precision of the length calculation obtained improves on those previously obtained by a single module. To better check the improvement of the use of two

**Table 2** Experimental results of the length measures obtained by a module as a function of the orientation angle

|  | 30° | | 45° | |
| --- | --- | --- | --- | --- |
|  | 10 m/s | 20 m/s | 10 m/s | 20 m/s |
| AVG (m) | 4.16 | 4.50 | 3.54 | 3.99 |
| STD (m) | 0.22 | 0.24 | 0.30 | 0.22 |
| Abs. error (m) | 0.46 | 0.80 | 0.16 | 0.29 |
| Rel. error (%) | 12.44 | 21.55 | 4.43 | 7.75 |

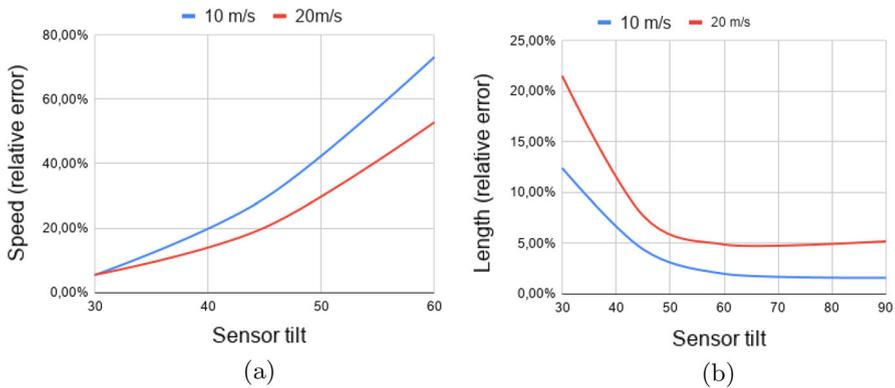

**Fig. 8** Comparative of the evolution of the speed errors **a** and length errors **b** according to the different angles used by the modules and interpolating the values for the intermediate angles

devices, on graphics 8a, b, the error comparative between devices (30 and 45°) with a single module and devices with two modules (30° and 90°).

Graph 8a shows the 60° obtained data, for comparative effects. The relative error of 60° is very high, so it is not included in the previous data charts. As it is shown, the relative error with the 30° angle is so similar. The difference between the speed errors increases as the angle increases. This means that the more parallel the angle is in the axis of the road, the less incidence of speed in error. In the case of the length calculation, using a single module obtains calculation errors above 5% in most cases. In the case of a module with the distance sensor at an angle of 30° with the axis of the road, the speed calculation is good, but the samples dedicated to the length are much fewer. This means that the length error becomes high despite the little speed error. An intermediate point is obtained in calculating the speed with a distance sensor at an angle of 45°, where both the rate and the length do not have a tolerable relative error.

However, when using two modules where the calculated speed by the first one has a low relative error, the length error is much smaller. Consequently, using a system where the best cases are combined implies a considerable improvement in calculating the two parameters, speed, and length.

The result obtained is similar in error to that expected using similar technologies. In [23], speed is obtained with accuracy from $\pm 0.2$ to $\pm 1$ m/s, whereas in our system, we obtain a speed error of 0.5 m/s, obtained from the relative error for a vehicle at a speed of $10 m/s$. However, in the system provided, we also obtain the vehicle's length from the speed provided by the 30° module and the data provided by the 90° module. This suggests that using more modules with different angle measurements allows for improved vehicle characterisation.

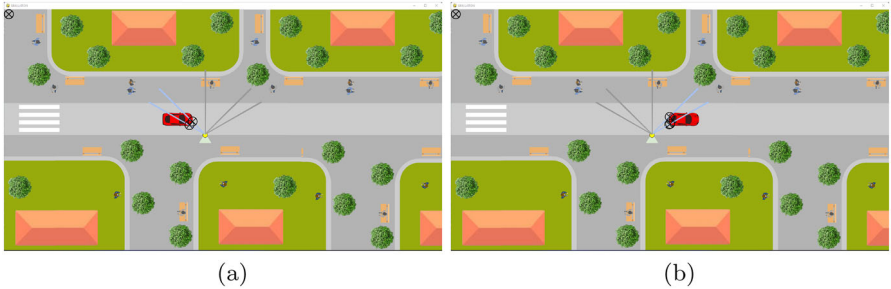

**Fig. 9** 20 m/s and a device composed of five modules with orientations of 30°, 45°, 90°, 135°, and 150°. In the left image, the vehicle is detected at the front, while in the right image the vehicle is detected at the rear

**Table 3** Experimental results of the length measurements obtained by the 90° module as a function of the speed provided by the 30° module

|  | 90° | |
| --- | --- | --- |
|  | 10 m/s | 20 m/s |
| AVG (m) | 3.76 | 3.89 |
| STD (m) | 0.15 | 0.21 |
| Abs. error (m) | 0.06 | 0.19 |
| Rel. error (%) | 1.61 | 5.22 |

### 3.3.3 Multiple module device

Previous experiments with the prototype have shown that the speed-accuracy depends on the angle at which the module orients the ultrasonic sensor. The last of the experiments presented consists of using a device with modules oriented at angles that provide relevant information. To test all the sensors in operation, the device has been simulated by modifying the traffic intersection simulator presented in [41]. This simulator is developed in Python, and the modified scenario and sensor simulation can be obtained from the author's repository [42]. The sensor is simulated using the signal obtained from the previously presented experimental device as the standard. This way, the simulated signal is similar to the experimental, so the response can be considered valid.

In order to check how the increase of modules reinforces the characterisation of the vehicles, a device with angles of 30°, 45°, 90°, 135° and 150° is simulated. In this way, in addition to detecting the front of the vehicle, the rear of the vehicle is also detected. The aim is to verify the vehicle speed better, as it can be obtained from sensors detecting the front of the vehicle as well as from sensors detecting the rear of the vehicle. The speeds of the vehicles are the same as those of the experimental prototype: 10 m/s and 20 m/s. Figure 9 shows what the simulator looks. Specifically, Fig. 9a shows how the simulated modules detect the front of the vehicle, while Fig. 9b shows the detection of the rear of the vehicle.

The simulated vehicle is similar to the vehicle used in the experiments. In the simulation, the signal does not need noise filtering, and only the CUSUM is simulated for detecting the relevant points and the derivative test for calculating the velocities. Table 4 shows the speeds obtained from the front side (30°, and 45°) and rear side (135°, and 150°), and compared with the Table 1, the relative error comparing the measures of these tables is 12.2%, quite similar of the Rel. error of each measure.

Table 5 shows the vehicle length obtained from each module without sharing data between them, and compared with the Table 2, the Relative Error comparing the measures of these tables is 10.1%, quite similar of the Rel. error of each measure.

**Table 4** Simulated results of the speed measures obtained by each module as a function of the orientation angle

|  | 30° | | 45° | | 135 | | 150 | |
| --- | --- | --- | --- | --- | --- | --- | --- | --- |
|  | 10 m/s | 20 m/s | 10 m/s | 20 m/s | 10 m/s | 20 m/s | 10 m/s | 20 m/s |
| AVG (m/s) | 10.09 | 19.99 | 9.87 | 19.95 | 9.32 | 19.08 | 10.04 | 19.38 |
| STD (m/s) | 1.60 | 2.39 | 1.41 | 1.33 | 1.29 | 2.34 | 1.51 | 2.14 |
| Abs. error (m/s) | 1.36 | 2.04 | 1.17 | 1.12 | 1.19 | 1.87 | 1.27 | 2.84 |
| Rel. error (%) | 13.62 | 10.21 | 11.69 | 5.58 | 11.92 | 9.36 | 12.67 | 9.18 |

**Table 5** Simulated results of the length measures obtained by each module of the device as a function of the orientation angle

|  | 30° | | 45° | | 135° | | 150 | |
| --- | --- | --- | --- | --- | --- | --- | --- | --- |
|  | 10 m/s | 20 m/s | 10 m/s | 20 m/s | 10 m/s | 20 m/s | 10 m/s | 20 m/s |
| AVG (m) | 3.80 | 3.46 | 3.73 | 3.49 | 3.49 | 3.30 | 3.71 | 3.23 |
| STD (m) | 0.02 | 0.03 | 0.01 | 0.03 | 0.03 | 0.05 | 0.01 | 0.06 |
| Abs. error (m) | 0.10 | 0.24 | 0.03 | 0.21 | 0.21 | 0.39 | 0.01 | 0.47 |
| Rel. error (%) | 2.63 | 6.36 | 0.81 | 5.71 | 5.65 | 10.66 | 0.26 | 12.69 |

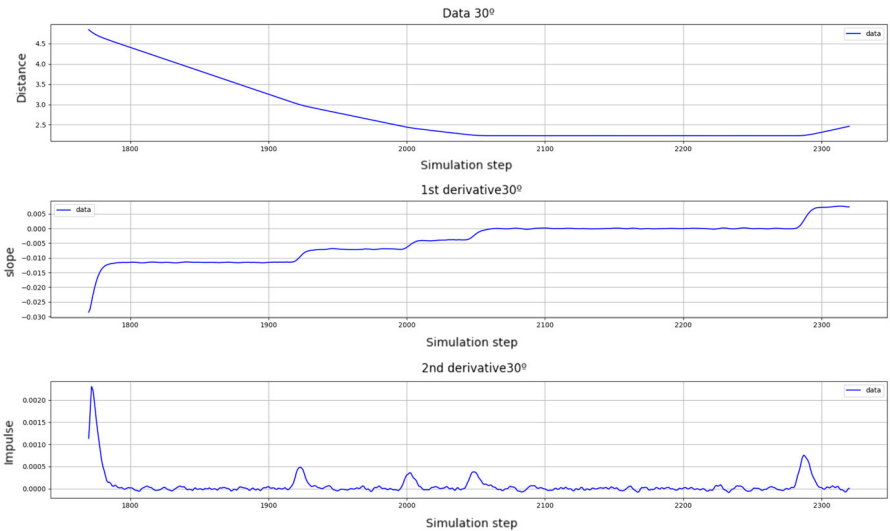

**Fig. 10** Simulation. The first graph shows the filtered signal. The second graph shows the first derivative The third graph shows the second derivative. The latter shows the changes in the signal trend, indicating possible changes in the vehicle's perimeter

Comparing this two tables with the experimental data (Tables 1 and 2), since the simulation has a greater capacity to generate and store data than the embedded device, the results of the derivative test of the 30° modules are shown in Fig. 10. The rest of the angles simulated have a similar profile.

**Table 6** Results of the length measurements obtained by the 90° module as a function of the speed provided by the 30° and 150° modules. All results were obtained by simulation

|  | 90° | |
| --- | --- | --- |
|  | 10 m/s | 20 m/s |
| AVG (m) | 3.84 | 3.50 |
| STD (m) | 0.02 | 0.02 |
| Abs. error (m) | 0.15 | 0.20 |
| Rel. error (%) | 3.98 | 5.47 |

In the figure, the upper graph shows the distance obtained to the front of the vehicle on the Y-axis and the samples obtained on the X-axis. The middle graph shows the first cumulative derivative of the distance. This first derivative shows the speed at which the edge of the vehicle is approaching. Finally, the bottom graph shows the second derivative, where the impulse shows the variations of the detected vehicle edge. This impulse is used to locate significant changes in the speed of the vehicle profile, e.g. the difference between the sides or, depending on the position, e.g. a rear-view mirror shows a very high impulse. From this derivative test, the speed of the vehicle's front end is deduced, as well as the significant changes in the different parts.

Finally, because the 90° module has the start and end of the side of the vehicle well located, it uses the data from the sensors detecting the front and rear of the vehicle to optimise the measurement made. Table 6 shows the results obtained in the simulation, and compared with Table 3, the Relative error comparing the measures of these tables is 4.1%, quite similar to the relative error of each measurement.

With all this, knowing where the vehicle's speed would be best calculated is possible. This case would be the front of the vehicle. The length is then obtained with the time difference between the detection of the right rear headlamp and the right front headlamp of the vehicle, with the previously obtained speed.

## 4 Conclusions

This article presents an implementation of a collaborative modular device system. The device consists of similar modules that provide distances to detect a vehicle and characterise length and speed. The novelty presented is using single modules to measure, with low accuracy, vehicle speed and length. Other contribution is how a decoupled but collaborative system can significantly improve the accuracy of measurements. It goes from a relative error of 5% at best to less than 1% when two modules collaborate. Although far from the errors obtained by camera vision systems, these errors are a good result due to the advantages they have in terms of cost and simplicity of computation.

Globally, the results may extrapolate to cybernetic systems. They are systems where the devices could collaborate from a basic level, like the module, to a higher level, like it is the information given from the cloud.

The next step consists of spreading various devices along a road. These devices can characterise the movement of vehicles. It is possible to detect both infractions and risk situations from the characterisation carried out. Among the infractions from a third device, it is possible to detect an acceleration that could lead to exceeding the established limits.

These limits are especially sensitive in the non-urban road environments where the featured devices are located.

It is easy to detect possible invasions of lanes in the opposite direction or sudden stops of vehicles due to the distance data regarding risk situations or accidents. To provide the functionalities mentioned above, more than two devices must collaborate. In this case, the device net conforms to an I2I system where it is necessary, in various points, to provide the information to another system that stores and analyses the history of each vehicle. It is also possible to share more advanced information between devices. Besides the speed and length, it is possible to send data relative to the vehicle behaviour, like speeding or excessive sensor spacing that may indicate an encroachment on other lanes.

**Acknowledgements** This work was by the Spanish Science and Innovation Ministry: CICYT project PRESE-CREL: "Models and platforms for predictable, secure and reliable industrial information technology systems" PID2021-124502OB-C41. Funding for open access charge: Open Access funding provided thanks to the CRUE-CSIC agreement with Springer Nature.

## Declarations

**Conflict of interest** The authors have no conflict of interest to declare. All co-authors have seen and agree with the manuscript's contents, and there is no financial interest to report.

## References

1. Broy M, Cengarle MV, Geisberger E (2012) Cyber-physical systems: imminent challenges. In: Monterey workshop, pp 1–28. Springer
2. Wolf W (2009) Cyber-physical systems. IEEE Ann Hist Comput 42(03):88–89
3. Serpanos D (2018) The cyber-physical systems revolution. Computer 51(3):70–73
4. Hichri Y, Dahi S, Fathallah H (2021) Candidate architectures for emerging IOV: a survey and comparative study. Des Autom Embed Syst 25(4):237–263
5. Canas V, Garcia A, de Las-Morenas J, Blanco J (2019) Modular and reconfigurable platform as new philosophy for the development of updatable vehicular electronics. Rev Iberoam de Automática e Inform Industr 16(2):200–211
6. Asselin-Miller N, Biedka M, Gibson G, Kirsch F, Hill N, White B, Uddin K (2016) Study on the deployment of c-its in Europe: final report. Rep for DG MOVE MOVE/C 3:2014–794
7. Corli A, Malaguti L (2021) Wave fronts in traffic flows and crowds dynamics. In: Cicognani M, Del Santo D, Parmeggiani A, Reissig M (eds) Anomalies in partial differential equations. Springer, Cham, pp 167–189
8. Maity S, Bhattacharyya A, Singh PK, Kumar M, Sarkar R (2022) Last decade in vehicle detection and classification: a comprehensive survey. Arch Comput Methods Eng 28:1–38
9. Sun Z, Bebis G, Miller R (2004) On-road vehicle detection using optical sensors: a review. In: Proceedings. The 7th International IEEE Conference on Intelligent Transportation Systems (IEEE Cat. No. 04TH8749), pp 585–590. IEEE
10. Li W, Li H, Wu Q, Chen X, Ngan KN (2019) Simultaneously detecting and counting dense vehicles from drone images. IEEE Trans Industr Electron 66(12):9651–9662

11. Messoussi O, Magalhães FGd, Lamarre F, Perreault F, Sogoba I, Bilodeau G-A, Nicolescu G (2021) Vehicle detection and tracking from surveillance cameras in Urban scenes. In: International symposium on visual computing, pp 191–202. Springer
12. Lozano Dominguez JM, Mateo Sanguino TJ (2019) Review on v2x, i2x, and p2x communications and their applications: a comprehensive analysis over time. Sensors 19(12):2756
13. Qian X, Hao L (2020) Performance analysis of cooperative sensing over time-correlated Rayleigh channels in vehicular environments. Electronics 9(6):1004
14. Hidalgo C, Marcano M, Fernandez G, Perez J (2020) Cooperative maneuvers applied to automated vehicles in real and virtual environments. Rev Iberoam de Autom e Inform industr 17(1):56–65
15. Scaglia G, Serrano ME, Albertos Pérez P (2020) Control de trayectorias basado en álgebra lineal. Rev Iberoam de Autom e Inform industr 17(4):344–353
16. Chen Z, Wu C, Huang Z, Lyu N, Hu Z, Zhong M, Cheng Y, Ran B (2017) Dangerous driving behavior detection using video-extracted vehicle trajectory histograms. J Intell Transport Syst 21(5):409–421
17. Celik T, Kusetogullari H (2009) Solar-powered automated road surveillance system for speed violation detection. IEEE Trans Industr Electron 57(9):3216–3227
18. de Oliveira M, Teixeira R, Sousa R, Tavares Gonçalves EJ, et al (2021) An agent-based simulation to explore communication in a system to control urban traffic with smart traffic lights
19. Adarsh S, Kaleemuddin SM, Bose D, Ramachandran K (2016) Performance comparison of infrared and ultrasonic sensors for obstacles of different materials in vehicle/robot navigation applications. IOP Conf Ser Mater Sci Eng 149:012141
20. Appiah O, Quayson E, Opoku E (2020) Ultrasonic sensor based traffic information acquisition system; a cheaper alternative for its application in developing countries. Sci Afr 9:00487
21. Matsuo T, Kaneko Y, Matano, M (1999) Introduction of intelligent vehicle detection sensors. In: Proceedings 199 IEEE/IEEJ/JSAI international conference on intelligent transportation systems (Cat. No. 99TH8383), pp 709–713. IEEE
22. Jo Y, Choi J, Jung I (2014) Traffic information acquisition system with ultrasonic sensors in wireless sensor networks. Int J Distrib Sens Netw 10(5):961073
23. Li SE, Li G, Yu J, Liu C, Cheng B, Wang J, Li K (2018) Kalman filter-based tracking of moving objects using linear ultrasonic sensor array for road vehicles. Mech Syst Signal Process 98:173–189
24. Jeon S, Kwon E, Jung I (2014) Traffic measurement on multiple drive lanes with wireless ultrasonic sensors. Sensors 14(12):22891–22906
25. Stiawan R, Kusumadjati A, Aminah NS, Djamal M, Viridi S (2019) An ultrasonic sensor system for vehicle detection application. J Phys Conf Ser 1204:012017
26. Marwedel P (2021) Embedded system design: embedded systems foundations of cyber-physical systems, and the internet of things. Springer, New York
27. Swanson DC (2011) Signal Processing for intelligent sensor systems with MATLAB. CRC Press, Boca Raton
28. Bae I, Ji U (2019) Outlier detection and smoothing process for water level data measured by ultrasonic sensor in stream flows. Water 11(5):951
29. Zhao H, Wang Z (2011) Motion measurement using inertial sensors, ultrasonic sensors, and magnetometers with extended Kalman filter for data fusion. IEEE Sens J 12(5):943–953
30. Ting J-A, Theodorou E, Schaal S (2007) Learning an outlier-robust Kalman filter. Eur Conf Mach Learn 25:748–756
31. Johnson NL (1961) A simple theoretical approach to cumulative sum control charts. J Am Stat Assoc 56(296):835–840
32. Pierre R.Bertrand, Hadouni D (2015) Change point detection by filtered derivative with p-value : choice of the extra-parameters. Working paper or preprint. https://hal.science/hal-01240885
33. Găşpăresc G, Gontean A (2014) Performance evaluation of ultrasonic sensors accuracy in distance measurement. In: 2014 11th International symposium on electronics and telecommunications (ISETC), pp 1–4. IEEE
34. Tsitsiklis JN (1988) Decentralized detection by a large number of sensors. Math Control Signals Syst 1(2):167–182
35. Koldobsky A (1998) Second derivative test for intersection bodies. Adv Math 136(1):15–25
36. Lee M, Newman RE, Latchman HA, Katar S, Yonge L (2003) Homeplug 1.0 powerline communication lans-protocol description and performance results. Int J Commun Syst 16(5):447–473
37. Poza-Lujan JL, Uribe-Chavert P, Sáenz-Peñafiel J-J, Posadas-Yagüe J-L (2021) Distributing and processing data from the edge. a case study with ultrasound sensor modules. Int Symp Distrib Comput Artif Intell 95:190–199

38. Marko D, Hrubỳ D (2020) Distance measuring in vineyard row using ultrasonic and optical sensors. In: Proceeding of 22 Nd international conference of Young scientists. Praha: Česká Zemědělská univerzita, pp 194–204
39. Semiconductors P (2000) The i2c-bus specification. Philips Semicond 9397(750):00954
40. Huh J-H, Seo K (2017) An indoor location-based control system using Bluetooth beacons for IoT systems. Sensors 17(12):2917
41. Gandhi MM, Solanki DS, Daptardar RS, Baloorkar NS (2020) Smart control of traffic light using artificial intelligence. In:2020 5th IEEE international conference on recent advances and innovations in engineering (ICRAIE), pp 1–6. IEEE
42. Uribe Chavert P, Gandhi MM (2022) Vehicle speed and length detector with ultrasound sensor and different angles. https://doi.org/10.5281/zenodo.7215268.'github.com/puch18ou/'